\documentclass[a4paper,10pt]{amsart}
\usepackage{graphicx}
\usepackage{adjustbox}
\usepackage{threeparttable}
\usepackage{tablefootnote}
\usepackage{cite}
\usepackage{algorithmic}
\usepackage{indentfirst,csquotes}

\usepackage{xcolor,paralist,hyperref,etoolbox}
\newtheorem{theorem}{Theorem}[]
\newtheorem{definition}[theorem]{Definition}

\hypersetup{ colorlinks=true, linkcolor=black, filecolor=black, urlcolor=black }

\begin{document}
\title{A Systematic Security Analysis for Path-based Traceability Systems in RFID-Enabled Supply Chains} 
\author[F. Heikamp]{Fokke Heikamp}
\author[L. Pan]{Lei Pan}
\author[R. Doss]{Robin Doss}
\author[R. Trujillo-Rasua]{Rolando Trujillo-Rasua}
\author[S. Ruj]{Sushmita Ruj}
\date{\today}
\email{f.heikamp@deakin.edu.au}

\let\thefootnote\relax

\begin{abstract}
Traceability systems have become prevalent in supply chains because of the rapid development of RFID and IoT technologies. 
These systems facilitate product recall and mitigate problems such as counterfeiting, tampering, and theft by tracking the manufacturing and distribution life-cycle of a product.
Therefore, traceability systems are a defense mechanism against supply chain attacks and, consequently, have become a target for attackers to circumvent. 
For example, a counterfeiter may change the trace of a fake product for the trace of an authentic product, fooling the system into accepting a counterfeit product as legit and thereby giving a false sense of security.  

This systematic analysis starts with the observation that security requirements in existing traceability solutions are often unstructured or incomplete, leaving critical vulnerabilities unaddressed.
We synthesized the properties of current state-of-the-art traceability solutions within a single security framework that allows us to analyze and compare their security claims. 
Using this framework, we objectively compared the security of $17$ traceability solutions and identified several weaknesses and vulnerabilities. 
This article reports on these flaws, the methodology we used to identify them, and the first security evaluation of traceability solutions on a large scale.
\end{abstract} 
\footnote{\emph{Corresponding author}: f.heikamp@deakin.edu.au}

\maketitle

\section{Introduction}
\label{sec:introduction}
Counterfeiting, theft, tampering, and product recalls are persistent challenges in modern supply chains~\cite{hassijaSurveySupplyChain2021}. 
Traceability systems are designed to address these issues by creating and maintaining a comprehensive, immutable, and searchable record of a product’s history.
To achieve this goal, traceability systems employ technologies such as RFID and IoT.
Academic literature describes many traceability systems for RFID-enabled supply chains~\cite{biSecureEfficientTwoParty2023,buEveryStepYou2018,burbridgeSupplyChainControl2010,islamIntegratingBlockchainSupply2022a,louSESCFSecureEfficient2021a,mamunSupAUTHNewApproach2018,qiDESwordIncentivizedVerifiable2022a,qianLightweightPathAuthentication2018,raySecureObjectTracking2016,wangRFChainDecentralizedCredible2023,wangEfficientTagPath2016,yangReSCRFIDEnabledSolution2018}.
Real-world industry-grade traceability systems are widely deployed, such as `Food Trust' by Walmart and IBM~\cite{WalmartIBMTraceability}.
Recently, the FDA has required traceability records for certain foods~\cite{FSMAFinalRule}. 
The Whitehouse issued an executive order demanding that developers provide a software bill of materials for their code~\cite{ExecutiveOrderImproving}.
Current trends suggest that traceability systems are going to play an even more crucial role moving forward.

Security is an important property of traceability systems.
However, most traceability solutions do not meet or partially meet the critical security requirements. 
The security requirements in existing work are either unstructured or incomplete~\cite{agrawalTraceabilitySovereignDistributed2006,dengResearchTraceabilityScheme2021,islamIntegratingBlockchainSupply2022a,mondalBlockchainInspiredRFIDBased2019a,uddinBlockchainMedledgerHyperledger2021}, creating an urgent need to secure traceability solutions~\cite{gs1GS1GlobalTraceability,qianLightweightPathAuthentication2018,wangEfficientTagPath2016}.
The impact can be enormous if an adversary can fabricate, modify, or delete records in a traceability system.
For example, a malicious supply chain participant could deny participation by deleting certain records, or a counterfeiter could sell his counterfeits as authentic by including a valid trace.
The difficulty in providing a comprehensive security analysis is partly because of the broad definition of traceability.
Therefore, we propose a more focused and concise definition of traceability.
We encapsulate this definition within a security framework, that, to the best of our knowledge, enables the first large-scale security evaluation of traceability solutions.

\noindent \emph{Contributions.} In this systematic analysis, we inspected the security of several traceability systems for RFID-enabled supply chains. 
We observed that all systems, implicitly or explicitly, defined traceability in terms of paths through a supply chain. The path returned by a secure traceability solution is expected to satisfy two properties. 
First, all readers on the path were visited physically, which is a spatial property.
Second, the readers were visited in the correct order, which is a temporal property.
These expectations have been used in path authentication papers~\cite{blassTrackerSecurityPrivacy2011,buEveryStepYou2018,elkhiyaouiCheckerOnsiteChecking2012,mamunSupAUTHNewApproach2018,ouafiPathcheckerRFIDApplication2009,wangEfficientTagPath2016}, yet have been neglected by many traceability solutions. 
Our main goal is to determine whether existing traceability solutions fulfil these expectations in the presence of an attacker, shedding light on the difficulty of the task and the limitations of the existing solutions. 
We do this by designing a security framework capable of expressing path-based attacks on traceability systems, providing a minimalist list of path-based functional and security requirements, and evaluating $17$ prominent traceability solutions. 
 
In terms of security flaws and vulnerabilities, our most relevant findings are a linking attack on RF-Chain~\cite{wangRFChainDecentralizedCredible2023} that violates path privacy, a path authorization attack on the solution proposed by Burbridge and Soppera~\cite{burbridgeSupplyChainControl2010}, an out-of-order attack on Ray et al.~\cite{raySecureObjectTracking2016}, an out-of-order attack on Tracker~\cite{blassTrackerSecurityPrivacy2011}, and a key-disclosure attack on ReSC~\cite{yangReSCRFIDEnabledSolution2018}. 

Our results show that all traceability systems are vulnerable to rerouting attacks with passive readers. 
In addition, a significant number of solutions are vulnerable to unauthorized path attacks. 
Finally, we noticed that a significant number of solutions do not require the readers to be visited in order~\cite{blassTrackerSecurityPrivacy2011,qianLightweightRFIDSecurity2016a,caiDistributedPathAuthentication2012a}.
The solutions that include path authorization do not explain how to distribute authorization policies securely.

\noindent \emph{Structure of the article.}
Section \ref{sec:relatedwork} presents the related work. 
Section \ref{sec:framework} introduces RFID-enabled supply chains, our generic traceability model, a taxonomy of path-based attacks, adversary models, and security properties. 
In Section \ref{sec:methodology}, we discuss the methodology of our framework. 
Section \ref{sec:results} presents the results of applying the proposed evaluation framework to numerous traceability systems.
We also discuss the key takeaways and limitations of our approach.
In Section \ref{sec:conclusion}, we conclude our paper and discuss future work.

\section{Related Work}
\label{sec:relatedwork}


The timing of a systematic analysis of supply chain traceability is particularly relevant given the recent regulatory requirements on traceability, the fragility of the supply chain exposed by the COVID-19 pandemic, and the large number of traceability solutions whose security remains unassessed. 
The challenge is that only a few security frameworks exist for traceability systems. 
For example, Cai et al.~\cite{caiNewFrameworkPrivacy2012} developed a privacy framework for path authentication solutions.
They focus only on RIFD tags and readers and do not consider back-end servers.
Their framework only considers privacy, whereas our framework includes security properties.
Syed et al.~\cite{syedTraceabilitySupplyChains2022} presented a threat analysis for a generic traceability system using the STRIDE framework.
They divided a traceability solution into four layers: data carrier, data capture, data sharing, and application.
Their threat analysis is detailed, but informal, complicating a formal comparison between traceability solutions.
Hassija et al.~\cite{hassijaSurveySupplyChain2021} conducted a survey of supply chain security.
They explore vulnerabilities in modern supply chains and present solutions, such as physically unclonable functions.
They focus on supply chains as a whole, whereas we concentrate on RFID-based traceability systems.
Gandino et al.~\cite{gandinoSecurityProtocolRFID2017} provided an RFID-based traceability solution using multi-entity encryption.
They provide a list of security requirements and conduct a structured informal security analysis. 
They listed attacks without providing a complete attack taxonomy. 
Their work focused on the RFID part and did not include back-ends. 
In contrast to their work, we use paths to formalize traceability systems and can prove the security properties more formally. 
Alzahrani et al.~\cite{alzahraniImprovedIoTRFIDEnabled2022} provided a traceability solution that can resist several attacks such as replay attacks and impersonation attacks. 
However, they did not provide a framework for the analysis. 

The absence of a simple and effective security framework for traceability systems may have played a role in the absence of a security analysis in many traceability solutions, including~\cite{contiFoodTraceabilityFruit2020,elshayebRFIDTechnologyZigbee2009,heSecureRFIDbasedTrack2008,uddinBlockchainMedledgerHyperledger2021}, and in the short and informal security analysis performed in \cite{burbridgeSupplyChainControl2010,Shi2012,islamIntegratingBlockchainSupply2022a,louSESCFSecureEfficient2021a,musamihBlockchainBasedApproachDrug2021}.
This approach contrasts with the approach followed by designers of path authentication protocols, 
such as Tracker \cite{blassTrackerSecurityPrivacy2011}, PathChecker \cite{ouafiPathcheckerRFIDApplication2009}, Checker \cite{elkhiyaouiCheckerOnsiteChecking2012}, StepAuth \cite{buEveryStepYou2018}, and SupAuth \cite{mamunSupAUTHNewApproach2018}, 
which attempt to provide cryptographic proofs of security.  
These studies are known for their formal security analysis, often using game-based security proofs and oracles.
However, path-authentication protocols concentrate on the data carrier and capture layers, resulting in an incomplete analysis of a supply chain traceability ecosystem.
That is, it is unclear whether the security of path authentication can be generalized to the security of supply chain traceability.
Our study aims to unify these properties in a single framework.
In other words, we wish to combine the formal precision of path authentication with the scope of traceability systems.

Table \ref{tab:related_work} summarizes the characteristics of the security analyses discussed in terms of four aspects: 
the presence of an \emph{evaluation framework} that can be used for the analysis of other designs, inclusion of path-based properties, coverage of all layers of the supply chain stack, such as the physical and application layer, and inclusion of an \emph{attack taxonomy}.

\begin{table*}[!ht]
    \tiny
    \centering
    \caption{An Overview of the Related Work}
    \label{tab:related_work}
    \begin{tabular}{|p{12em}|p{4.5em}|p{4.5em}|p{4.5em}|p{4.5em}|p{16em}|}
        \hline 
        \textbf{Article} & \textbf{Evaluation Framework} & \textbf{Path-Based Traceability} & \textbf{Complete Security Analysis} & \textbf{Attack Taxonomy} & \textbf{Description} \\
        \hline
        Cai et al.~\cite{caiNewFrameworkPrivacy2012} & O & \checkmark & X & X & Did only provide a framework for privacy on the data carrier and capture layer. \\
        Syed et al.~\cite{syedTraceabilitySupplyChains2022} & \checkmark & X & O & \checkmark & Does not use a path-based approach to traceability. Security analysis is informal and there is no case study. \\
        Gandino et al.~\cite{gandinoSecurityProtocolRFID2017} & \checkmark & X & X & O & Only focuses on RFID and not back-ends. Does not consider paths. \\
        Path Authentication~\cite{blassTrackerSecurityPrivacy2011,buEveryStepYou2018,elkhiyaouiCheckerOnsiteChecking2012,mamunSupAUTHNewApproach2018,ouafiPathcheckerRFIDApplication2009} & \checkmark & O & X & O & Does not include back-end in security analysis. Does not connect path authentication to traceability. \\
        Alzahrani et al.~\cite{alzahraniImprovedIoTRFIDEnabled2022} & X & X & \checkmark & O & Informal security analysis. Do not provide a security framework. \\
        Our Framework & \checkmark & \checkmark & O & \checkmark & Provides a security framework for path-based traceability. Includes an attack taxonomy. \\
        \hline
    \end{tabular}
\end{table*}

Although our systematic analysis of supply chain traceability is new, there are similar studies on different but related topics.
Avoine et al.~\cite{avoineFrameworkAnalyzingRFID2011} proposed a security framework for distance-bounding protocols.
They later used this framework to do a survey~\cite{avoineSecurityDistanceBoundingSurvey2019} of all distance-bounding protocols.
Our work is similar but focuses on supply chain traceability. 
Ladisa et al.~\cite{ladisaSoKTaxonomyAttacks2023} provided an attack taxonomy for open-source software supply chains.
Their main contribution is an attack tree for open-source software supply chains.
They use expert interviews to verify their taxonomy.
They mention traceability by including the software bill of materials (SBOM) in their safeguard table.
However, they did include path-based attacks in their taxonomy.
He and Zeadally~\cite{heAnalysisRFIDAuthentication2015} performed a systematic analysis of RFID authentication schemes using elliptic curves.
Maleki et al.~\cite{malekiSoKRFIDbasedClone2017} provided a systematization of knowledge (SoK) for clone detection mechanisms in RFID-enabled supply chains.

\section{Our Framework}
\label{sec:framework}
In this section, we introduce the components of a traceability system for RFID-enabled supply chains, such as readers, tags, and back-end servers.
Additionally, we define system behaviour as statements about the movement of tags and provide a taxonomy of path-based properties. 

\subsection{RFID-enabled Supply Chains}
\label{sec:framework-RFID-SC}

An RFID-enabled supply chain utilizes RFID technology to track and manage products across the entire supply chain.
Tags and readers are the essential components in such a system. 
Each RFID tag is uniquely assigned to a single product. 
Because of this one-to-one correspondence, our framework models only the RFID tag to represent the product.
However, to trace tags, some systems in the literature rely on back-end and data-sharing servers. 
We refer to those systems as \emph{online} systems because they require an active connection between readers, back-end servers, and (possibly) data-sharing servers to make statements regarding the movement of a tag. 
Examples of systems based on an active back-end server include~\cite{qiDESwordIncentivizedVerifiable2022a,raySecureObjectTracking2016,wangRFChainDecentralizedCredible2023}. 
Examples of systems based on active blockchain or EPCDS~\cite{gs1GS1GlobalTraceability} data sharing solutions are~\cite{islamIntegratingBlockchainSupply2022a,musamihBlockchainBasedApproachDrug2021,yangReSCRFIDEnabledSolution2018}. 
We shall refer to \emph{offline} systems as systems that can make traceability statements about an RFID tag by simply relying on the information stored in the tags' memory, such as in~\cite{blassTrackerSecurityPrivacy2011,buEveryStepYou2018,qianLightweightPathAuthentication2018,wangEfficientTagPath2016}.

\begin{figure}[!ht]
     \centering
     \includegraphics[width=0.6\columnwidth]{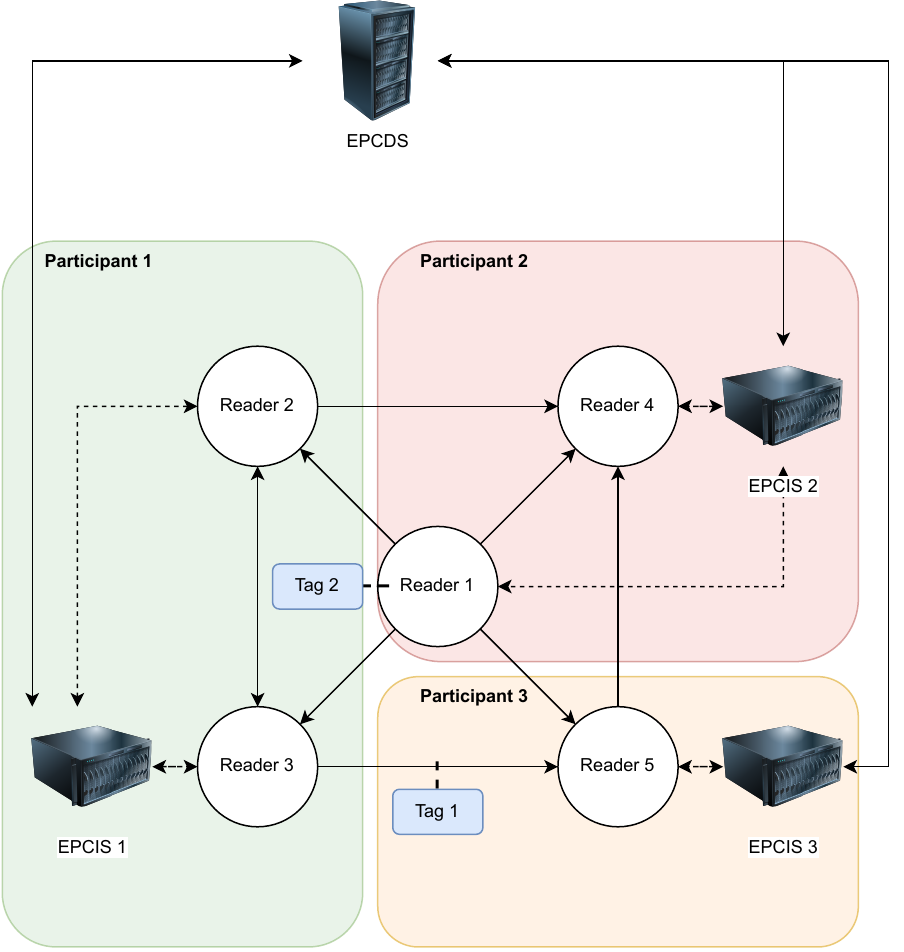}
     \caption{Our RFID-enabled Traceability Model}
     \label{fig:completemodel}
\end{figure}

To capture all components of a traceability system, including offline and online characteristics, we followed the GS1 EPCGlobal Standard~\cite{gs1GS1GlobalTraceability}, as shown in Figure~\ref{fig:completemodel}. 
Readers were assigned to participants, each of which had a back-end.
Back-ends have more capabilities than RFID readers in terms of data management capabilities, including data storage, computation, and connectivity. 
A data-sharing server $ds$ is assumed to exist to facilitate data sharing among all the readers. 
An entity not depicted in Figure \ref{fig:completemodel}, yet present in most traceability solutions, is the \emph{issuer}: a trusted entity that generates and distributes secret keys, initializes the internal state of tags and readers. 
We are now ready to formally define a traceability system. 

A traceability system is a tuple $(R, T, B, ds, \mathcal{I})$ where $R$ is a set of readers, $T$ is a set of tags, $B$ is a set of servers, $ds$ is a data-sharing server and $\mathcal{I}$ is the issuer. 
A \emph{path} in the system is any sequence of readers $r_1 \ldots r_n$.
Note that we do not consider the \emph{products} part of a supply chain traceability system. 
We can abstract away from products because, similar to most traceability solutions in the literature, we assume perfect binding between products and tags. 
In other words, we do not distinguish between tracing a tag and tracing a product.
The only exceptions to this approach were ReSC~\cite{yangReSCRFIDEnabledSolution2018}, Anandhi et al.~\cite{anandhiIoTEnabledRFID2019}, and DE-Sword~\cite{qiDESwordIncentivizedVerifiable2022a}. 


\subsection{Behavior of a traceability system}
\label{sec:framework-gts}

Although all traceability systems aim to track products, there is no standard method in the literature to specify traceability systems and their behavior. Most systems are described informally, whereas others would use mathematical rigor for the specification and analysis of their communication protocols only. 
Therefore, our task is to provide a common language that could serve as the Rosetta Stone for understanding traceability systems for RFID-enabled supply chains. 

\newcommand{\pset}{\mathop{\mbox{\fontfamily{msa}\selectfont P}}}

We define the behavior (also known as semantics) of a traceability system $T$ by the set of traces it produces, denoted $traces(T)$, where a trace is a sequence of events. Formally, given a universe of events $\mathcal{E}$, a trace is an element of $\mathcal{E}^*$ and the semantics of a traceability system is a subset of $\pset(\mathcal{E})$ where $\pset(\cdot)$ denotes the power set. 
We reserve the following events in $\mathcal{E}$: 
\begin{itemize}
    \item $\textsc{Move}(t, r)$ to denote the physical movement of a tag $t$ to the location of reader $r$.
    \item $\textsc{ValidPath}(t, r_1, \ldots, r_n)$ denotes that the sequence of readers $r_1 \cdots r_n$ is a valid path for $t$ to take. In general, this event is issued by a trusted entity, such as the issuer.  It can be used to establish physical constraints and expectations on the path that a tag should follow. 
    \item $\textsc{Path}(t, r_1, \ldots, r_n)$ to denote a statement made by an authorized verifier in the system, such as a reader or a backend server, stating that tag $t$ has gone (supposedly) through the path $r_1 \cdots r_n$. 
\end{itemize}

Our definition of behavior is generic enough to accommodate most (if not all) traceability systems, and, it can be easily mechanized by modeling a system as a labeled transition system. 
However, the latter is beyond the scope of this study. 
Our focus is on defining the correctness properties for traceability systems that use events with the meaning established above. 
We begin by defining the physical path taken by a tag based on an execution trace. 

\begin{definition}[Physical Path]
\label{def:physicalpath}
Let $\tau$ be a trace and $t$ a tag. 
The \emph{physical path} of $t$ in $\tau$, denoted as $path_{\tau}(t)$, is recursively defined as follows:
\begin{itemize}
    \item If $\tau = e_1$ and $e_1 = \textsc{Move}(t, r)$, then  $path_{\tau}(t) = r$, otherwise $path_{\tau}(t)$ is the empty sequence. 
    \item If $\tau = e_1 \cdots e_n$, let $r_1 \cdots r_m = path_{e_1 \cdots e_{n-1}}(t)$. If $e_n = \textsc{Move}(t, r)$ and $r_m \neq r$, then $path_{\tau}(t) = r_1 \cdots r_m r$, otherwise $path_{\tau}(t) = r_1 \cdots r_m$.   
\end{itemize}
\end{definition}

Note that the definition above regards tag looping around the same reader as a single movement. 
That is, our theory of movement accepts cycles but not loops. 


Next, we establish the main goals and properties of a traceability system. 
According to the GS1 standard,  traceability refers to the ability to track the history and location of a product~\cite{gs1GS1GlobalTraceability}.
We sought a more precise definition of traceability. 
Therefore, we used \emph{paths} as building blocks.

\begin{definition}[Path-based properties]
\label{def:gts}
Let $\tau = e_1 \cdots e_n$ be a trace with $e_i = \textsc{Path}(t, r_1, \ldots, r_n)$ for some $i \in \{1, \ldots, n\}$. Let $r_1' \cdots r_m' = path_{e_1 \cdots e_{i-1}}(t)$. 
We say that the path statement $e_i$  
is, 
\begin{itemize}
    \item \textbf{sound} if $\{r_1, \ldots, r_n\}$ is a proper subset of $\{r_1', \ldots, r_m'\}$, that is if all readers in the path event have indeed been visited by $t$. 
    \item \textbf{complete} if $\{r_1, \ldots, r_n\} = \{r_1', \ldots, r_m'\}$, i.e. if the path event correctly provides all readers visited by $t$.
    \item \textbf{sorted} if $r_1 \ldots r_n$ is a subsequence of $r_1' \cdots r_m'$, i.e., if the path event provides a (possibly partial) history of the readers visited by $t$ in the right order. 
\end{itemize}
We say that a traceability system is sound, complete, or sorted if all path events produced by the system are sound, complete, or sorted. 
\end{definition}

The properties above are related in the following ways: i) a sorted system is sound but not necessarily complete, and ii) a complete system is sound but not necessarily sorted. 
In general, we are interested in systems that are sorted and possibly complete. 
Nevertheless, we include soundness because, although weaker than sorted, it provides a correct set of readers visited by a tag. 
Moreover, as we will show later, some systems can only achieve soundness in the presence of an attacker.
For simplicity, we assume that a system satisfies \emph{path authentication} if it is both sound and sorted.
Similarly, we define a system that satisfies \emph{complete path authentication} if it is sound, sorted, and complete.

An orthogonal property of traceability systems that we consider is \emph{authorization}. 
Whereas path-based properties are statements about whether a path statement correctly provides the trajectory of a tag, an authorization property establishes whether a path statement satisfies the constraints imposed on the movement of tags.
For example, a system may establish that every product that arrives at $r$ should subsequently be processed by $r'$, hence all paths should ensure that $r$ is immediately followed by $r'$. 
In our language, these constraints can be modeled using $\textsc{ValidPath}(\cdot)$ events. 

\begin{definition}[Authorization]
\label{def:authorization}
Let $\tau = e_1 \cdots e_n$ be a trace with $e_i = \textsc{Path}(t, r_1, \ldots, r_n)$ for some $i \in \{1, \ldots, n\}$. 
We say that path event $e_i$ is \emph{authorized} if there exists $j < i$ such that $e_j = \textsc{ValidPath}(t, r_1 \ldots r_n r_{n+1} \cdots r_m)$ with $n \leq m$. 
\end{definition}

In other words, a path is authorized if it is the prefix of a valid path. Again, we could require authorized paths to be complete, that is enforce $n = m$. 
However, based on our literature study, we find such restriction unnecessary.  

Another relevant property that often appears in the literature~\cite{blassTrackerSecurityPrivacy2011,caiNewFrameworkPrivacy2012}, which is somewhat associated with physical paths, is \emph{privacy}. 
Privacy claims are difficult to express using traces or paths because they depend on information that an adversary can obtain.
Therefore, we informally defined path privacy outside our framework.
We leave a formal integration of privacy properties into our framework for future work.
We used the path privacy property defined by Cai et al.~\cite{caiNewFrameworkPrivacy2012}. 
Path privacy is a combination of \emph{step unlinkability} and \emph{tag unlinkability}~\cite{blassTrackerSecurityPrivacy2011,wangTwolevelPathAuthentication2012}.
The tag unlinkability property ensures the integrity of a tag's content without leaking identifiable information.
Step unlinkability ensures that the tag does not leak path information to unauthorized participants.

The informal definition for path privacy is given in Definition~\ref{def:pathprivacy}.

\begin{definition}[Path Privacy]
\label{def:pp}
Path privacy holds if adversary $Adv$ cannot:
\begin{itemize}
    \item Determine if two tags $t_1$ and $t_2$ have any steps in common
    \item Determine if two tags $t_1$ and $t_2$ are the same tag
\end{itemize}
\label{def:pathprivacy}
\end{definition}



Table \ref{tab:nomenclature} lists the terminology used in this study.

\begin{table}[!ht]
    \centering
    \caption{Terminology}
    \label{tab:nomenclature}
    \begin{tabular}{|p{8em}|p{26em}|}
        \hline
        \textbf{Symbol} & \textbf{Description}  \\
        \hline
         $GTS$ & Generic traceability system, defined by a tuple 
$(R, T, B, ds, \mathcal{I})$\\
         $R$ & the set of all readers \\ 
         $B$ & the set of all back-ends \\ 
         $T$ & the set of all tags \\
         $P$ & the set of all participants \\
         $ds$ & the server used for data sharing \\
         $r_i^p$ & indicates that reader $r_i \in R$ belongs to participant $p \in P$ \\
         $pp_t$ & indicates the physical path (absolute truth) of the tag $t$ \\
         $cp_t$ & indicates the claimed path of tag $t$ returned by a traceability solution \\
         $VP_t$ & the set of valid paths for a given tag $t$ \\
         \hline
    \end{tabular}
\end{table}

\subsection{Path-based Attack Taxonomy}
\label{sec:framework-attacks}
\begin{figure*}[!ht]
    \centering
    \includegraphics[width=0.9\textwidth]{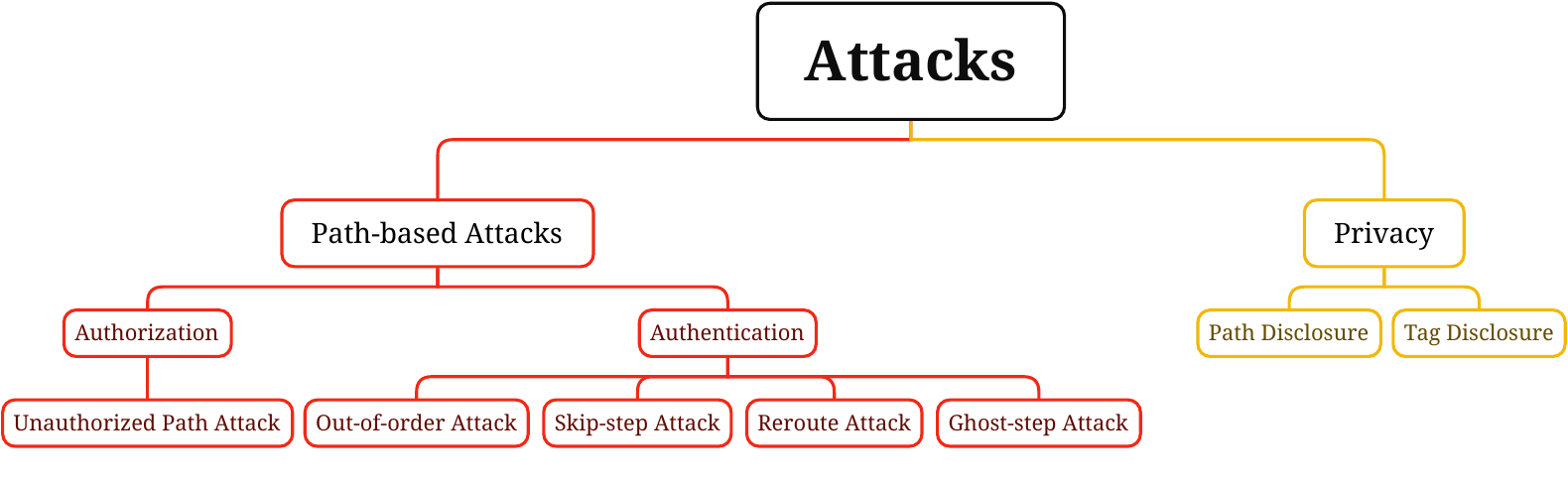}
    \caption{Taxonomy for Path-based Attacks}
    \label{fig:taxonomy}
\end{figure*}

Multiple types of anomalies can be identified from a path-based perspective~\cite{buUnveilingMysteryInternet2020}.
First, a traceability system can report a reader that has never been visited.
Second, a traceability system could fail to return a reader that was visited.
Third, a traceability system could incorrectly record the order of readers.
Lastly, it could return a path that the tag was not allowed to follow.
We categorized the attacks into three types: authentication, authorization, and privacy attacks.
Privacy attacks cannot be demonstrated from a path-based perspective.
However, we included these attacks for completeness.
The complete taxonomy is illustrated in Figure \ref{fig:taxonomy}.

\begin{table}[!ht]
    \centering
    \caption{Path-based Attacks.}
    \label{tab:attacks}
    \begin{tabular}{|p{8em}|p{6em}|p{6em}|p{16em}|}
        \hline
        \textbf{Name} & \textbf{Physical Path} & \textbf{Claimed Path} & \textbf{Description} \\
        \hline
        \emph{Out-of-order Attack} & $(r_1, r_2, r_3)$ & $(r_1, r_3, r_2)$ & All steps are there but not in the right order \\
        \emph{Skip-step Attack} & $(r_1, r_2, r_3)$ & $(r_1, r_3)$ & Reader $r_2$ has not been documented \\
        \emph{Reroute Attack} & $(r_1, r_2, r_x, r_3)$ & $(r_1, r_2, r_3)$ & Rerouted through $r_x$ \\
        \emph{Ghost-step Attack} & $(r_1, r_2, r_3)$ & $(r_1, r_2, r_x, r_3)$ & Step $r_x$ is documented but was never visited \\
        \hline
    \end{tabular}
\end{table}


Each attack has different impacts on supply chains:
\begin{itemize}
    \item An \emph{Out-of-order Attack} occurs if the \emph{claimed path} returned by the traceability system is in the incorrect order. 
    For example, consider a product that follows the paths ``production'', ``usage'', and ``quality control''.  
    Quality control took place after being used, indicating a participant might have tried to change the order in the traceability system. 
    Systems that do not satisfy the sorted property are vulnerable to this type of attack.
    \item In a \emph{Skip-step attack}, the traceability system fails to document valid steps. 
    An adversary may attempt this attack when it attempts to deny its involvement.
    Systems that are do not satisfy the completeness property are vulnerable to this attack.
    \item A \emph{Reroute attack} is similar to a Skip attack, except for a subtle difference. 
    A skip attack fails to document but is still on an authorized path. 
    Conversely, a reroute attack passes through a node that is not expected.
    \item A \emph{ghost-step attack} occurs when a traceability system documents a visited reader that has not been visited at all.
    It could incriminate honest participants. Systems that do not satisfy the soundness property are vulnerable to this attack.
    \item Finally, the tag can follow an \emph{unauthorized path}. 
    Detecting unauthorized paths is essential for identifying errors at an early stage.
\end{itemize}

It is worth mentioning that Table ~\ref{tab:attacks} reuses terminology found in the literature.
Our work mapped these attacks to path-based properties.

\subsection{Adversary Models}
\label{sec:framework-adv-models}

The security properties of traceability systems should be assessed against a realistic adversary model. 
Regarding communication capabilities, most traceability solutions consider the Dolev-Yao adversary model \cite{dolevSecurityPublicKey1983}, which can inject, modify, read, and delete all messages sent to the network. 
Regarding compromising capabilities, the literature considers two types of adversaries: one that can compromise tags but not readers~\cite{mamunSupAUTHNewApproach2018,qianLightweightPathAuthentication2018,wangEfficientTagPath2016,yangReSCRFIDEnabledSolution2018}, and another that can compromise both tags and readers. 
Back-end and data-sharing servers are considered trusted and reliable~\cite{buEveryStepYou2018,qiDESwordIncentivizedVerifiable2022a}.

Thus, we develop two adversary models to match these approaches: the \emph{tag compromise adversary mode}, $Adv_{T}$ for short, and the \emph{reader compromise adversary model}, $Adv_{R}$ for short.  
Both adversary models are Dolev-Yao models with the capacity to read and write the content of the RFID tags. In addition, the $Adv_{R}$ model can also compromise readers. 

Security properties cannot hold in the presence of an attacker that compromises the entire system. 
Readers making security claims are generally considered trusted, as in~\cite{zhangEndtoEndTraceabilityICs2020}. We leave these considerations to the specific properties that a system targets.    
We want to keep the definitions sufficiently generic to apply to most traceability systems while preserving a good level of detail.



In our adversary models, we do not consider physical attacks that involve physical interactions with tags.
Physical attacks have been described in several studies ~\cite{avoineRFIDTraceabilityMultilayer2005,ranasingheSecurityPrivacySolutions2004,weisSecurityPrivacyAspects2004}.
Owing to the limited capabilities of RFID tags, it is difficult to protect against physical attacks without significantly increasing the cost.
This increases the complexity of the proposed model.
Our results demonstrated the effectiveness of this a path-based approach.


\section{Methodology}
\label{sec:methodology}
Our framework enables an objective comparison of multiple systems.
To test this framework, we identified and analyzed $17$ relevant traceability systems.
We use the following query to search exhaustively for traceability systems: (``RFID'' OR ``radio frequency identifier'') AND (``supply chain'') AND (``traceability'' OR ``object tracking'' OR ``product tracing'' OR ``provenance'' OR ``path authentication'') AND (``secure'' OR ``security'' OR ``threats''). 
We used the default search settings and searched through abstracts, titles, and keywords. 
We searched for papers in well-known databases such as IEEE xplore, Scopus, and the ACM digital library.
We used three criteria to determine relevancy.
First, a traceability solution must be implemented in the selected study.
Some studies claim to provide a traceability solution but they implement something else.
For example, Anandhi et al.~\cite{anandhiIoTEnabledRFID2019} provided a mutual authentication scheme for RFID tags but did not build a traceability system.
Second, we require that they use RFID technology. 
Finally, the authors must make security claims about their traceability system.
We excluded some of our initial search results.
Some studies did not provide any security claims~\cite{heSecureRFIDbasedTrack2008}.
Other studies did not use RFID technology~\cite{louSESCFSecureEfficient2021a} or implement a traceability solution~\cite{alzahraniImprovedIoTRFIDEnabled2022,anandhiAuthenticationProtocolTrack2020a,anandhiIoTEnabledRFID2019,mamunLightweightMultipartyAuthentication2021}.
Finally, some papers are extensions of earlier studies~\cite{raySecureObjectTracking2015,raySecureObjectTracking2016,mamunRFIDPathAuthentication2014,mamunSupAUTHNewApproach2018}.
In this case, we analyze the most recent system.

Our workflow for analyzing a traceability system is as follows.
First, we translated the target system into a generic traceability system by mapping the components of the target to a system $(R, T, B, ds, \mathcal{I})$.
If a system consists of components beyond those in our model, we omit them.
We documented these components in an informal manner.
For example, a traceability system can use watermarks on a product. 
In this case, our model can verify security properties but not watermarks. 
If a traceability system does not use specific components in our RFID model (such as a back-end or data-sharing server), we exclude them.
The only important distinction from a functional perspective is that the traceability system requires online access.
We call a traceability system offline if it can function using only the information stored on the RFID tag.
We call a traceability system online if it requires access to a back-end or data-sharing server.
We included this distinction in our results table and labeled it as an \emph{architecture}.

After translating the components of the target system into our components, we must describe the communication between them.
We accomplish this by summarizing the system using message sequence charts.
A message sequence chart shows the interactions between different components.
In this chart, we have an abstract notion of cryptography and must make assumptions about messages, if the paper is ambiguous.
We used the events introduced in Section~\ref{sec:framework-gts} to describe the system.

After describing the target system, we conduct a security analysis.
The security analysis consists of two steps.
The first step is a generic security analysis, in which we highlight the weaknesses and vulnerabilities outside the scope of our framework.
The second step is to apply the introduced security framework 
to assess whether 
the target system resists the attacks described in Section~\ref{sec:framework-attacks}.
We demonstrate resistance against attacks using the path-based and privacy properties mentioned in Section~\ref{sec:framework-gts}.
We indicate whether the system holds under the $Adv_{T}$ or the $Adv_{R}$ models.


\begin{table*}[!htbp]

    \begin{adjustbox}{max width=\textwidth}
    \begin{threeparttable}[b]
        \caption{Analyzed Solutions}
        \label{tab:analysis}
        \centering
        \begin{tabular}{|p{10em}|p{8em}|p{5em}|p{4em}|p{4em}|p{5em}|p{4em}|p{14em}|}
            \hline
            \textbf{Name} & \textbf{Building Blocks} & \textbf{Architecture} & \textbf{Sorted and Sound} & \textbf{Sorted, Sound, and Complete} & \textbf{Authorization} & \textbf{Privacy} & \textbf{Description} \\
            \hline            
            Burbridge and Soppera~\cite{burbridgeSupplyChainControl2010} & Proxy re-signatures & Offline & $Adv_{T}$\tnote{4} & X & $Adv_{T}$\tnote{1} & X & An $Adv_{R}$ adversary can bypass authorization. Achieves no path privacy. Not explained how policies are established and distributed.   \\
            RF-Chain~\cite{wangRFChainDecentralizedCredible2023} & PKI, blockchain & Online & $Adv_{T}$\tnote{2} & X & X & $Adv_{T}$\tnote{1} & Vulnerable to hash length extension. Vulnerable to a privacy linking attack. No path authorization. \\
            ReSC~\cite{yangReSCRFIDEnabledSolution2018} & PKI, Physically Uncloneable Function & Online & $Adv_{T}$\tnote{1} & X & $Adv_{R}$ & X & Vulnerable to impersonation attacks. Stores session keys in plain text on the tag.  \\
            DE-Sword~\cite{qiDESwordIncentivizedVerifiable2022a} & ZK-EDB, Game Theory & Online & $Adv_{R}$ & X & X & $Adv_{R}$ & No threat model for external adversaries. Assumes participants cannot distinguish between good and bad products. Querying strictly after distribution. \\
            SupAuth~\cite{mamunSupAUTHNewApproach2018} & Homomorphic Message Authenticators\cite{catalanoPracticalHomomorphicMessage2018} & Offline & $Adv_{T}$ & X & $Adv_{T}$ & X & Uses same key on all readers. Does not explain how path information is exchanged with the checkpoint. Possible out-of-order attack. \\
            SecTTS~\cite{Shi2012} & EPCGlobal Network & Online & $Adv_{T}$ & X & X & $Adv_{T}$\tnote{3} & Lack of details in description. Vulnerable to linking, replay, and impersonation attacks. Security depends on the number of fake EPCs. \\
            Two-level~\cite{wangTwolevelPathAuthentication2012} & PKI, HIBE Trees, Batch Verification & Online & X & X & X & $Adv_{T}$ & Uses fully trusted readers. Does not store paths. Ambiguity in path selection. HIBE tree has inherent weaknesses. \\ 
            End-to-End Traceability IC~\cite{zhangEndtoEndTraceabilityICs2020} & Ring Oscillator, PKI & Offline & $Adv_{R}$ & X & X & X & Data on tag is readable to everyone. Scalability issues. \\
            Cai et al.~(Ordered Multi-Signatures)~\cite{caiDistributedPathAuthentication2012a} & Polynomial Path Encoding, Shamir Secret-sharing, Ordered Multi-signature & Offline & $Adv_{T}$ & X & $Adv_{T}$ & X & Ordered multi-signature is a relatively untested building block. No privacy under our adversary model. Needs to store a lot of valid paths. Communication protocol with trusted back-end not specified. \\
            StepAuth~\cite{buEveryStepYou2018} & Nested Encryption, Hybrid Encryption & Offline & $Adv_{R}$ & X & $Adv_{R}$ & $Adv_{R}$ & No explicit protection against replay attacks. Only works with a single static path. Growing tag secret. \\
            Secure object tracking~\cite{raySecureObjectTracking2015,raySecureObjectTracking2016} & Physically Unclonable Function & Offline & $Adv_{T}$\tnote{1} & X & $Adv_{T}$\tnote{1} & X & Readers can be visited in any order, invalidating path authentication and authorization and leading to an Out-of-order Attack. Not possible to implement on low-end RFID tags. \\
            Islam et al.~\cite{islamIntegratingBlockchainSupply2022a} & Physically Unclonable Function & Online & $Adv_{T}$ & X & X & X & Uses a custom blockchain. Does not offer any privacy or path authorization. \\
            PathChecker~\cite{ouafiPathcheckerRFIDApplication2009} & PKI, PRF & Offline & $Adv_{T}$ & X & $Adv_{T}$ & X & Requires tag computation. Does not offer privacy. \\
            Tracker~\cite{blassTrackerSecurityPrivacy2011} & Polynomial Path Encoding & Offline & $Adv_{T}$\tnote{1} & X & $Adv_{T}$ & $Adv_{T}$ & Can only verify that a valid path has been taken, but cannot tell which path has been taken. Vulnerable to out-of-order attack.  \\
            Cai et al.~\cite{caiNewFrameworkPrivacy2012} & PKI & Offline & $Adv_{T}$\tnote{1}  & X & $Adv_{T}$ & $Adv_{T}$ & More efficient extension of Tracker. Shares the same weaknesses. \\
            Qian et al.~\cite{qianLightweightPathAuthentication2018} & PKI & Offline & $Adv_{T}$\tnote{1} & X & $Adv_{T}$ & $Adv_{T}$ & Do not provide path completeness. Extension of Tracker. \\
            Checker~\cite{elkhiyaouiCheckerOnsiteChecking2012} & Hash, Polynomial Path Encoding & Offline & $Adv_{T}$ & X & $Adv_{R}$ & $Adv_{R}$ &  Significant overhead for key distribution. Does not explain how valid paths are distributed. Solves the out-of-order attack that is present in Tracker. \\
            \hline
        \end{tabular}
        \begin{tablenotes}
            \item [1] Attack found
            \item [2] Weakness found
            \item [3] Depends on the amount of fake data
            \item [4] Only if there is only one valid path
        \end{tablenotes}
    \end{threeparttable}
    \end{adjustbox}
\end{table*}

\section{Results}
\label{sec:results}

Table~\ref{tab:analysis} summarizes our analysis results for $17$ traceability systems.
Each entry in the table documents the functional and security properties of the traceability solution. 
Functionally, we list the architecture (online or offline) and building blocks of the traceability solution. 
For security properties, we show whether the path and privacy properties are satisfied. 
We use $X$ to indicate that the security property is not satisfied.
Similarly, we use $Adv_{T}$ or $Adv_{R}$ to indicate that the security property holds under an $Adv_{T}$ adversary or $Adv_{R}$ adversary respectively.
We describe any meaningful results, such as weaknesses or attacks, in the column ``description''.

We identified several types of attacks and weaknesses.
For example, RF-Chain~\cite{wangRFChainDecentralizedCredible2023} has an attack on path privacy, Ray et al.~\cite{raySecureObjectTracking2016} is vulnerable to an out-of-order attack, Tracker~\cite{blassTrackerSecurityPrivacy2011} is also vulnerable to an out-of-order attack, ReSC~\cite{yangReSCRFIDEnabledSolution2018} is not resistant to impersonation attacks, and Burbridge and Soppera~\cite{burbridgeSupplyChainControl2010} did not satisfy completeness under the $Adv_{R}$ adversary.

\subsection{Summary of individual analyses}
 
\textbf{Burbridge and Soppera}~\cite{burbridgeSupplyChainControl2010} developed a traceability solution based on proxy re-signatures~\cite{atenieseProxyResignaturesNew2005,blazeDivertibleProtocolsAtomic1998}.
It utilizes an online architecture, so that it consists of readers, tags, back-ends, and a data-sharing server.
A proxy re-signature scheme~\cite{atenieseProxyResignaturesNew2005} transforms a signature for $a$ into a signature for $b$, but cannot generate signatures for either $a$ or $b$.
The supply chain controller(SCC) initializes every participant with shipping policies, receiving policies, key pairs, and re-signing keys.
These policies determine the routing of the tags.
Each participant checked whether they could receive goods according to their receiving policy.
Subsequently, it verifies the signature and looks up the next step according to the shipping policy.

Initially, their scheme had a key explosion problem because it required a unique key pair per entry in the policy table.
They solved this problem by using the same key for each reader and relying on the receiver for verification.
However, this introduces a weakness, in which an $Adv_{R}$ adversary can violate path authorization.
To show how this works, consider that tag $t_1$ has to follow path $(r_a, r_b, r_c, r_d, r_e)$, while tag $t_2$ has to follow path $(r_a, r_b, r_d, r_e)$.
This translates into two $\textsc{ValidPath}$ events: $\textsc{ValidPath}(t_1, r_a, r_b, r_c, r_d, r_e)$ and $\textsc{ValidPath}(t_2, r_a, r_b, r_d, r_e)$. 
If both $r_b$ and $r_d$ are dishonest, they can send $t_1$ through the path of $t_2$, bypassing participant $r_c$.
Thus, we obtain the following trace for $t_1$: 
\begin{multline}
tr_{t_1} = \textsc{Move}(t_1, r_a) \rightarrow \textsc{Move}(t_1, r_b)\rightarrow \textsc{Move}(t_1, r_d) \rightarrow \\
\textsc{Move}(t_1, r_e) \rightarrow \textsc{Path}(t_1, r_a, r_b, r_c, r_d, r_e)
\end{multline}

This trace shows that our authorization property is invalidated.
Having a different re-signature key for every tag solves this problem, but this causes a key explosion, that limits its scalability.
The authors did not provide an efficient solution for this problem.
Path authorization holds under the $Adv_{T}$ adversary model because readers are trusted in that adversary model.
Their scheme cannot guarantee completeness and soundness because it does not make a path claim.
Burbridge and Soppera did not explain how to establish these shipping policies.
The authors did not consider path privacy because the tag identifiers were sent in plain text.

We noticed that related work, analyzing Burbridge and Soppera, wrongly states that the scheme requires a trusted third party to compute the re-signatures~\cite{elkhiyaouiCheckerOnsiteChecking2012,raySecureObjectTracking2016}.
This is not the case because the re-signature is calculated by the participants, not a trusted third party.
Moreover, Bu and Li~\cite{buEveryStepYou2018} stated that the checkpoint only verifies at the path end, which is incorrect because every step can perform this check.

\textbf{Wang et al.}~\cite{wangRFChainDecentralizedCredible2023} proposed RF-Chain, a product authentication system that protects against counterfeiting.
RF-Chain~\cite{wangRFChainDecentralizedCredible2023} uses online and offline secrets.
The offline secret, stored on a UHF RFID tag, is defined as:
\begin{align*} 
a_0 &=  H(ID||f||pwd||r) \\ 
a_i &=  Sig_{v_i}(a_{i-1})
\end{align*}

The online secret, stored on the blockchain, uses the offline secret and is defined as follows:
\begin{align*} 
b_i &= a_{i-1}\oplus H(h_i),\;where\;h_i=ID||f||pwd||r||i
\end{align*}

\begin{figure}[!htb]
    \centering
    \includegraphics[width=0.95\columnwidth]{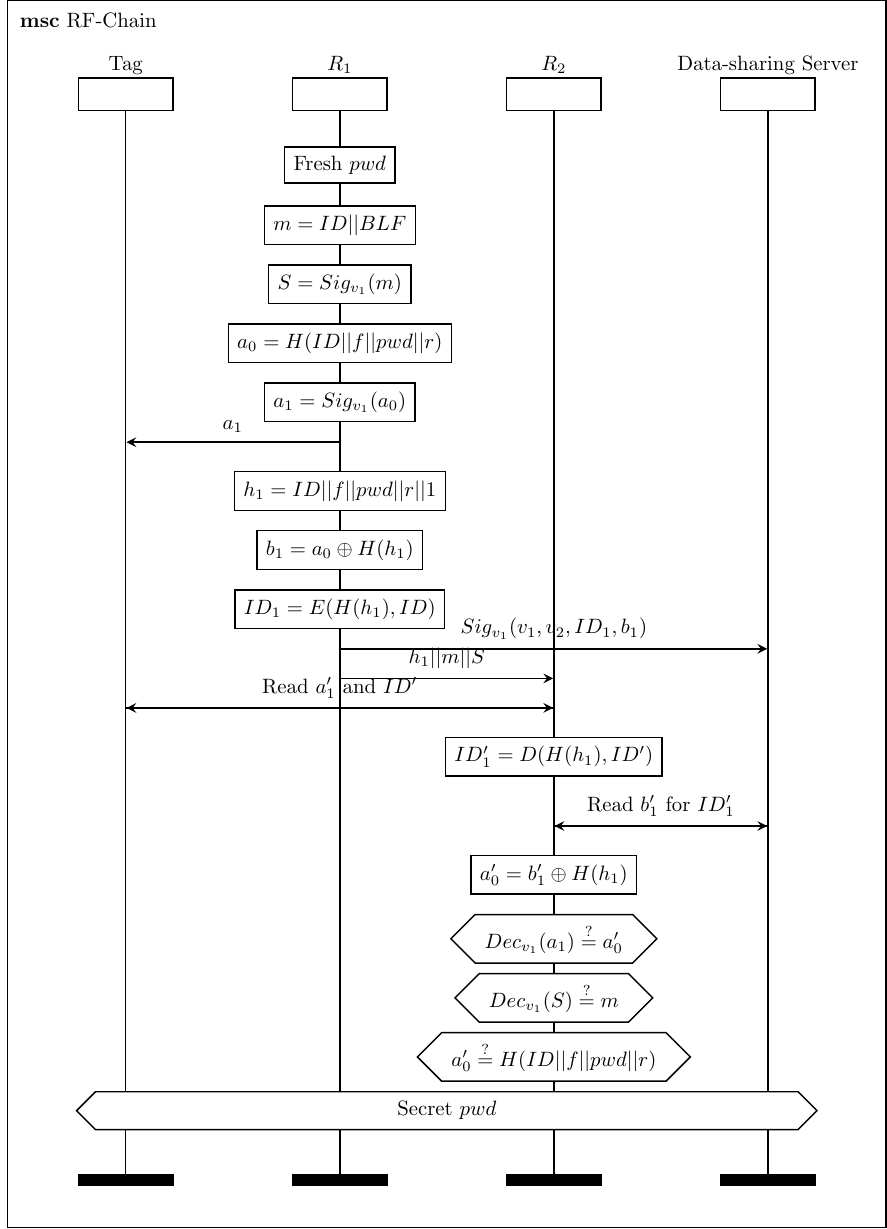}
    \caption{A single step of RF-chain}
    \label{fig:rf-chain}
\end{figure}

A message sequence chart for the interaction between the first and second readers is shown in Figure \ref{fig:rf-chain}.
All subsequent steps are the same, except that they use different values for $a$, $b$, or $h$.
Moreover, the values for $pwd$ and message $m$ remain the same for every tag. 
The last step is slightly different because its validity has to be checked.

The records on the blockchain are stored using a pseudo-identity.
The authors claimed that this pseudo-identity prevents linking attacks.
However, they reuse the hash value $H(h_i)$, which allows any adversary to link records together.
An attacker who can read the tag ID at least once can retrieve all the past information for this tag from the blockchain.
Because path privacy is not a path-related property we use the terminology from the original paper.
The attack works as follows:
\begin{itemize}
    \item Reading the $a_i$ value, the adversary can remove the signature to obtain $a_{i-1}$.
    \item The adversary retrieves all data from the blockchain to get the following list $((ID_1, b_1), \dots, (ID_n, b_n))$.
    \item The goal is to determine which $b$ belongs to $a_{i-1}$.
    \item For every $b_j, j \in 1\dots n$, we calculate possible key $H(h_x) = a_{i-1} \oplus b_j$.
    \item We tested whether $H(h_x)$ is a valid key by encrypting the tag with this key $ID_x = E(H(h_x), ID)$.
    \item If $ID_x = ID_1$, we found the correct key and correct $b$.
\end{itemize}
This process is repeated for all $j$, where $j<i-1$ to learn all values of $b$.
This attack invalidates Section 5.1.5 of their security analysis and is also a violation of path privacy.
Without this vulnerability, path privacy would hold only under the $Adv_{T}$ adversary model.
If a reader is compromised, an adversary can learn $H(h_i)$ for any $i$, meaning that they can link all the data on the blockchain because the pseudo-identity is encrypted with $Enc(H(h_i), ID)$.

We noticed a construction in which they used the hash $H(h_i)$ as an encryption key for $ID_i = E(H(h_i), ID)$.
Every $h_i$ is defined as $h_i = ID||f||pwd||r||i$.
This construction is vulnerable to hash length extension attacks.
Since adversaries can predict the input hash value, they can try to generate $h_j = ID||f||pwd||r||j$, for $j>i$.
However, this is not possible because of the padding.
Nonetheless, it is advisable to change its structure.
Their solution does not satisfy the path completeness property.
RF-Chain also does not satisfy the path authorization property because the authors did not consider this requirement.

\textbf{ReSC} is a traceability solution for IoT supply chains.
They targeted two problems: theft and counterfeiting.
More specifically, they focused on split attacks using a neighborhood attestation protocol.
IoT devices are used tTo achieve this.
In our model, we do not model the product itself; therefore, the IoT device itself is beyond the scope of this study.
Therefore, we can only analyze a subset of their solution.

Their architecture used tags, readers, and a central database.
This can be modeled using the complete model, as shown in Figure \ref{fig:completemodel}.
Because the authors do not use any back-ends, we model the back-end servers as proxies between the readers and the database.
Therefore, we have the components $(R, T, B, ds, \mathcal{I})$.

Their system has three phases: registration, tracing, and neighbor attestation.
During registration, the IoT device, tag, and back-end are initialized.
First, a CCID is generated using a physically uncloneable function (PUF) on an IoT device.
The tag also generates an identifier $ID$.
A tuple $(CCID, ID)$ is sent to the back-end server using a trusted channel. 
The tag gets initialized as $TID||P||k_1||sig_1||index_1||ts_1||\dots||k_n||sig_n||index_n||ts_n|$, where $TID$ is the tag ID, $P$ is a pointer pointing to the location of the next signature.
For all $i \in 1\dots n$: $k_i$ are the session keys, $sig_i$ are the signatures, $index_i$ represents the indices, and $ts_i$ are the timestamps.
Depending on where in the path we are, the signature, index, and timestamp might not have been added yet.
The tag is initialized with all session keys $k_i$.
The session keys are generated as $Enc_{k_r}(TID)$, where $k_r$ is a shared key between the reader and back-end.
The authors of ReSC define a tag trace as valid if all signatures are in place on the tag.

\begin{figure}[!htb]
    \centering
    \includegraphics[width=0.95\columnwidth]{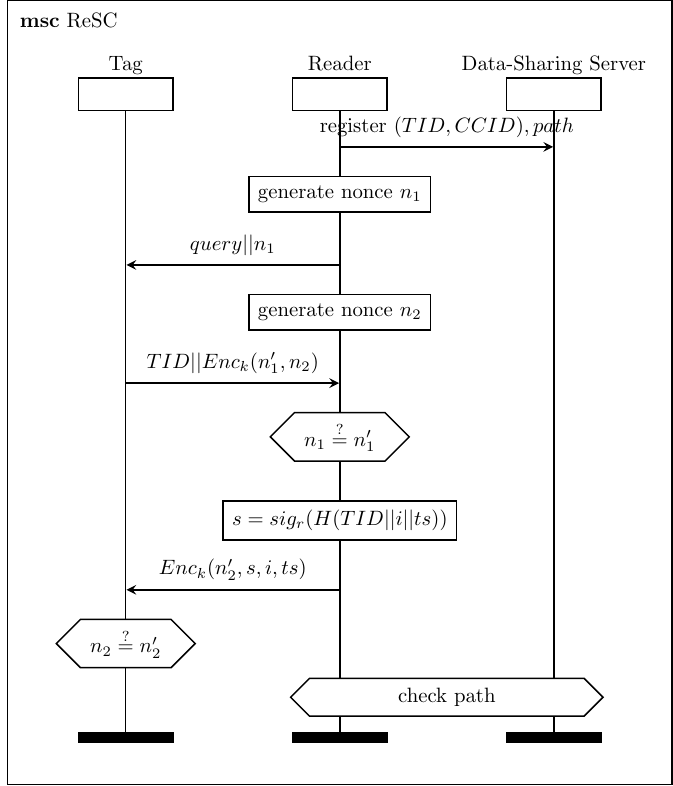}
    \caption{ReSC}
    \label{fig:resc}
\end{figure}

Resc defines a mutual authentication scheme between the reader and tag.
A message sequence chart for a single step of their protocol is shown in Figure~\ref{fig:resc}.

The authors specify that readers cannot be compromised, meaning the system should be analyzed in the $Adv_T$ adversary model.
Under this adversary model, path authentication and authorization are satisfied.
By initializing the session keys $(k_1, \dots, k_n)$ on the tag they make a $\textsc{ValidPath}(t, r_1, \dots, r_n)$, where each key $k_i$ is associated with reader $r_i$.
If the system makes a path claim $\textsc{Path}(t_1, r_1, \dots, r_n)$ it can be verified if the signatures on the tag correspond to readers $(r_1, \dots, r_n)$.
The order must also be correct because the timestamps are included and the readers are trusted.
Unauthorized parties cannot access the tag contents because there is a mutual authentication protocol.

However, this schemehas scalability and feasibility limitations.
RFID tags have limited resources.
According to the authors, the session keys take up $128$ bits, indices take up $20$ bits, and timestamps take $23$ bits.
The authors did not mention signature size.
If ECDSA signing is used with a $256$ bit curve, then the signature is $512$ bits.
Therefore, leaving the size of the tag identifier and pointer out of scope, the storage required for a path of length $n$ is $n * (128 + 20 + 23 + 512)$.
The most common RFID tags have approximately $512$ bits of memory.
In addition, most RFID tags cannot perform complex operations such as encryption or hashing.
Mutual authentication in RFID systems is an active research area, where the main challenge is to do it in a manner that does not require too many resources.

The RFID tag stores all the session keys.
Although it is true that the session keys never leave the tag unencrypted, it is unfeasible to expect that this information cannot be extracted.
RFID tags are inexpensive and simple electronic circuits that are unlikely to have complex security mechanisms.
In addition, if there is a dishonest reader, it can only read out all session keys.
This allows for impersonation attacks in the $Adv_R$ adversary model.

It does not achieve path privacy; the signatures are there to see for everyone who reads the tag.
The authors did not build systems for this purpose.

\textbf{SupAuth}~\cite{mamunSupAUTHNewApproach2018} is a path authentication solution that uses homomorphic encryption for security and privacy. 
The building blocks they use are called homomorphic message authenticators~\cite{catalanoPracticalHomomorphicMessage2018,gennaroFullyHomomorphicMessage2013}. 
A homomorphic message authenticator allows a server to perform computations on authenticated data without knowing the secret key.
In this case, a computation means that it can evaluate arithmetic circuits.
The server can convince other participants that it applies the computations correctly without disclosing the underlying data.
The components of SupAuth are readers, tags, manufacturer (issuer), and checkpoints.
Checkpoints are special types of reader that can perform path checks.

In their scheme, all readers share the secret key $sk$, whereas the tag serves as the server and only knows the encryption key $ek$.
Each reader and tag in the system are represented by a polynomial of degree 1 ($y(z) = y_0 + y_1z$).
When a reader receives a tag, it first authenticates it using its secret key $sk$ to obtain an authenticator.
It forwards the authenticator to the tag, which evaluates the tag over the arithmetic circuit.
The path is checked at the checkpoint.
The checkpoint can accomplish this because it maintains (directly or indirectly) a list of valid paths $\{\textsc{ValidPath}(t, R_1), \dots\, \textsc{ValidPath}(t, R_n)\}$, where $R_i$ is a subset of readers for $i \in 1\dots n$.

Functionally, this study has four limitations.
First, the key distribution was one key per tag.
Therefore, for every new tag, a new secret key must be distributed to all readers in the path which complicates the key management process. 
Second, they assume that the checkpoint knows which path the tag should take, but the authors do not describe how to synchronize this information between the manager and checkpoint.
Third, readers in between cannot verify the path taken because the checkpoint is strictly at the end. 
Finally, the tag must perform complicated calculations, which is unfeasible for simple RFID tags.

One weakness of their scheme is that all readers share a secret key $sk$, which means that SupAuth cannot function in our $Adv_{R}$ adversary model.
A dishonest reader can misuse the secret key to forge a valid path.
Soundness does hold because given a path claim $\textsc{Path}(t, r_1, \dots, r_n)$, tag $t$ must have gone through all readers by design.
It is unclear whether their solution is \emph{sorted}.
The evaluation of the circuit can likely be performed in any order, and the same result can be derived.
However, a formal proof must be provided.
This is left for future work.
Path authorization holds under the $Adv_{T}$ adversary model.
The authors did not consider path completeness or path privacy in their solution.

\textbf{SecTTS}~\cite{Shi2012} implements a privacy-friendly relay model based on the EPCglobal network standard.
There is a one-on-one match between the components of their system and our system.
The EPCDS can be matched with our data-sharing server, and the EPCISes function as back-ends.
Their solution needs to be online because it requires a data-sharing server.

\begin{figure*}[!htb]
    \centering
    \includegraphics[width=0.95\textwidth]{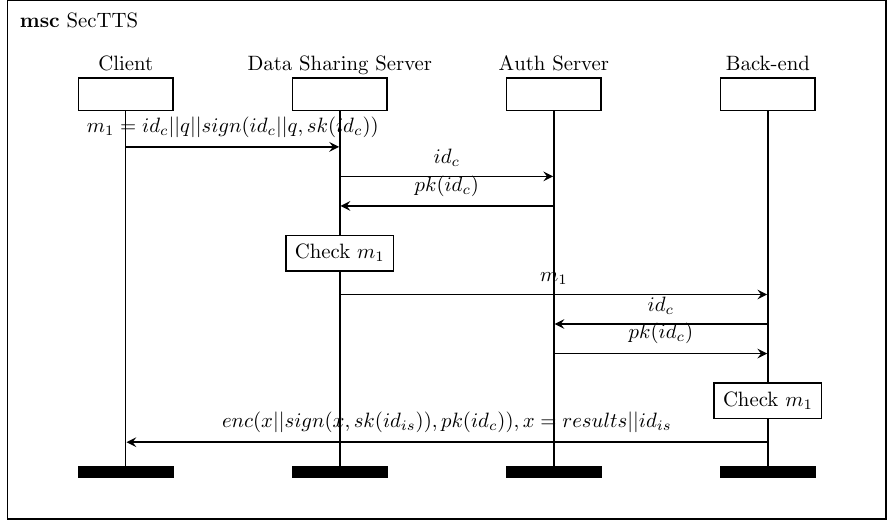}
    \caption{SecTTS}
    \label{fig:sectts}
\end{figure*}

Their solution uses set-based relay policies.
These policies can be either positive (allow) or negative (prohibit).
Each back-end publishes two lists of electronic product codes (EPCs) to the data-sharing server.
The first list describes all products the back-end handles (allow). 
The second list describes all products that they do not want to handle (prohibit).
Users can query a data-sharing server by providing an EPC.
The data-sharing server verifies whether any policy matches.
The query is forwarded to any back-end with an allow policy for this EPC.
Based on the system description of the authors, we reconstructed that the querying phase must resemble the message sequence chart in Figure~\ref{fig:sectts}.

The main security goal of the authors was to prevent passive attackers and the data-sharing server from learning which back-end handles which products.
The authors added fake entries to facilitate this.
Because of these fake entries, the data-sharing server cannot detect the back-end that processes a specific product.
SecTTS provides no resistance to the $Adv_{R}$ adversary because all participants are assumed to be honest.

One functional limitation is that the privacy of their scheme depends on the number of fake entries in the system.
This design decreases the performance.
The authors did not describe the number of fake entries that needed to be generated.
In addition, the authors did not explain how these fake entries were generated.
Their description does not protect against replay or impersonation attacks because the communication is signed but not encrypted.
We also noticed that an attacker could differentiate fake entries by observing query responses.
If a back-end receives a query for a fake EPC, it does not respond. 
A passive attacker can recognize fake queries by observing whether the query receives a response.
The messages to the data-sharing server and back-ends are not encrypted; therefore, the content can be read.
Their solution was sound and ordered under the $Adv_{T}$ adversary model.
However their system did not satisfy completeness.

\textbf{StepAuth}~\cite{buEveryStepYou2018} uses nested and hybrid encryption to encode a static path on a tag.
They used a trusted manager (Issuer) to set up a path secret on the tag.
Given a path $p=(r_1, \dots, r_n)$, the path secret is initialized by encrypting a message with the public key of every reader.
Therefore, a tag that must follow this path is initialized with $secret^1 = m, Sig_{k^{sec}}(m)$ where $m = Enc_{k_1^e}(k_1^s), Enc_{k_1^s}(1, 2, secret^2)$ and $secret^2$ are defined similarly.
This recursion continues until $secret^n$.
The secret defines a valid path $\textsc{ValidPath}(t, r_1, \dots, r_n)$.
Their solution works offline, because it does not depend on a back-end or data-sharing server.

At every step, the reader checks whether the secret can be decrypted.
If they can, they will peel off the layer and update the path secret on the tag.
This forces the system to generate a trace $tr_t = \textsc{Move}(t, r_1) \rightarrow \textsc{Move}(t, r_2) \rightarrow \dots \rightarrow \textsc{Move}(t, r_n) \rightarrow \textsc{Path}(t, r_1, \dots, r_n)$.
The manager signs the secret to protect against replay attacks and impersonation attacks.
So, the path is encoded on the tag and can only be reduced to the last message by peeling off all the intermediate layers, meaning their solution satisfied both path authentication (sound and sorted) and path authorization.
Furthermore, this works under the $Adv_{T}$ and $Adv_{R}$ adversary.
In their scheme, the tag identifier must remain unique.
Otherwise, it is vulnerable to replay attacks.
The tag content is encrypted, so path privacy is guaranteed as well.
The main limitation is that the path cannot change after initialization.

However, this scheme has several functional limitations. 
Unlike other schemes, such as Tracker~\cite{blassTrackerSecurityPrivacy2011} and Checker~\cite{elkhiyaouiCheckerOnsiteChecking2012}, tag secrets are growing.
Given that their solution requires $1024 + 896(l-1)$ bits of storage, where $l$ is the path length, at least 8kb is required to store a path of length $8$.
In our study, we rarely observed tags with more than $512$ bits of storage.
High-memory tags do exist, but in practice, they are rarely used.

\textbf{Ray et al.}~\cite{raySecureObjectTracking2016} proposed a secure object-tracking algorithm using a physically unclonable and pseudo-random function.
An uncloneable function prevents an adversary from cloning an RFID tag.
Their system contains three main components: the current owner $CO$, participants $P$, and tag $T$.
The $CO$ calculates a path code $PC = h(PID_1 \oplus \dots \oplus PID_n)$, where $h$ is a secure hash function and $PID_i$ is the identifier for participant $i$.
The $CO$ sets up a challenge value $c=h(PC \oplus RID_{CO})$, where $RID_{CO}$ is the identifier for the current owner.
The $CO$ then creates participant specific challenges $c_i=c \oplus PID_i$ and sens the challenges $\{c_1, \dots, c_n\}$ to the tag $t$.
Each participant uses its $c_i$ value to authenticate with the tag.

This scheme does not achieve path authentication or path authorization because it does not enforce the path to be visited in sequence.
Thus, in terms of traces it allows us to generate invalid traces such as:
\begin{multline}
tr_{invalid} = \textsc{Move}(t, r_x) \rightarrow \textsc{Move}(t, r_y) \rightarrow \textsc{Move}(t, r_z) \rightarrow\\
\textsc{Path}(t, r_y, r_x, r_z)
\end{multline}
This makes them vulnerable to \emph{out-of-order attacks}.

In addition, the communication between the reader and tag requires a secure channel.
Otherwise, the scheme will be vulnerable to replay attacks.
For example, reader $r_i$ sends challenge $c_i$ to the tag, which is used for authentication.
An adversary can intercept this value and block the request so that they can replay it at a later stage.

Their scheme is vulnerable to impersonation attacks by the $Adv_{T}$ and $Adv_{R}$ adversaries.
A dishonest participant can calculate any $c_i = (c \oplus PID_i \oplus PRF(n))$, for any $i \leq n$, since $c$, $PRF(n)$, and $PID_i$ are known.
An external adversary can capture a single $c_i$ and calculate $c$ because he knows $n$ and $pid_i$.
Thus, after observing one message, he can impersonate any reader in the path.
The authors do not explain how the pseudo-random functions are synchronised with each other, which complicates our analysis.

The path for each tag is pre-defined which makes the scheme only useful for scenarios where the complete path is known in advance.
Functionally, they require an RFID tag with a PUF and can execute XOR and hash functions.
These functions are normally not available for low-end inexpensive RFID tags.

\textbf{Tracker}~\cite{blassTrackerSecurityPrivacy2011} is a path authentication solution that uses polynomial encoding and homomorphic encryption.
The authors performed their security analysis using an honest but curious adversary model, which matches our $Adv_{T}$ adversary model.
They aim to provide path authentication, authorization, and privacy.

The main idea of Tracker~\cite{blassTrackerSecurityPrivacy2011} is that paths are stored as polynomials. 
This approach was first proposed for fault detection by Noubir et al.~\cite{NOUBIR1998405}. 
Their system consists of four components: readers, tags, an issuer, and a manager.
Therefore, no back-ends are involved.
The manager is a special type of reader that can verify the path of a tag.
Therefore, normal readers cannot verify the path of a tag, which is a functional limitation.

The setup of the scheme is as follows.
The issuer generates a key pair $(sk, pk)$, symmetric key $k$, generator $x_0$, and set of coefficients for each reader $\{a_1, \dots, a_n\}$.
Secret key $sk$ and symmetric key $k$ are shared only with the manager.
The values $a_i$ are shared only with the manager and reader $r_i$.
Generator $x_0$ is public knowledge.

Paths are encoded as polynomials in a finite field of order $q$, where $q$ is a prime number. 
The polynomial is defined as $Q_P(x) = a_0x^l + \sum_{i=1}^{l}a_ix^{l-i}$.
Given two distinct paths $P_1$ and $P_2$, they only have a probability of $1/q$ to evaluate to the same value. 
Thus, given a sufficiently large prime number, collisions are negligible. 
The tag secret is stored as $s_t^l = (E(ID), E(HMAC_k(ID)), E(\phi_{ID}(P)))$, where $\phi_{ID}(P) = HMAC_k(ID)\phi(P)$.
The authors use elliptic curve Elgamal encryption.
Owing to its homomorphic properties, readers can update the tag without decrypting it.
Every time reader $r_i$ sees a tag, it updates the tag content by applying a transition function that transitions $\phi_{ID}(P_{i-1})$ into $\phi_{ID}(P_{i})$. 

\begin{figure*}[!ht]
    \centering
    \includegraphics[width=0.95 \textwidth]{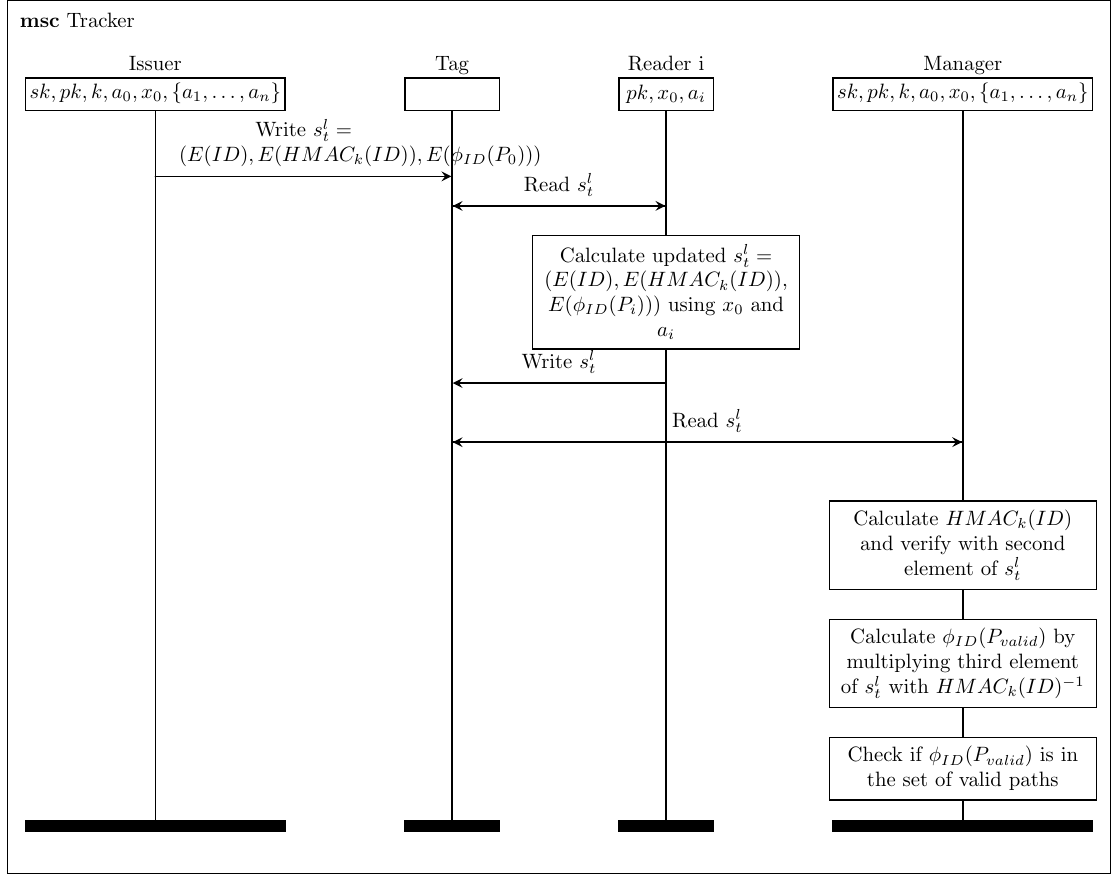}
    \caption{Tracker}
    \label{fig:tracker}
\end{figure*}

The manager is the only entity that can perform verification.
Because the manager knows the secret key $sk$ and all the coefficients $\{a_0, \dots, a_n\}$, it can create a list of polynomial evaluations coupled with their respective paths.
The manager also knows the secret key; therefore, it can decrypt and verify the path.
A high-level overview of the tracker protocol is shown in Figure~\ref{fig:tracker}.
Note that we only modeled one arbitrary reader $r_i$. 
In practice, there would be a chain of readers $(r_1, \dots, r_n)$.

The issuer maintains a set of valid paths 
\begin{equation*}
\{\textsc{ValidPath}(t, R_1), \dots\, \textsc{ValidPath}(t, R_n)\}
\end{equation*}
, where $R_i$ is a subset of readers for $i \in 1\dots n$.
Each movement of a tag $t$ between readers can be modeled as a $\textsc{Move}(t, r_i)$.
Their scheme satisfies \emph{soundness}: if a path claim $\textsc{Path}(t, r_1, \dots, r_n)$ is made, readers $r_1$ to $r_n$ update the tag.
Otherwise, the polynomial evaluation will not be the same. 

One difficulty is determining how these path claims are made.
In practice, the manager pre-computes the polynomial evaluation of valid paths.
It does not evaluate all the paths because it is computationally unfeasible.
Thus, if a tag has followed a non-valid path, the manager will detect it, but it cannot detect which path has been followed.
Meaning soundness can only be determined if the path is authorized because not all path evaluations are stored.
The authors did not describe the establishment of a set of valid paths.

\textbf{Cai et al.}~\cite{caiNewFrameworkPrivacy2012} proposed a new path authentication scheme that focused on path privacy.
They achieved this by using pseudo-random functions and ElGamal encryption.
The manager is the only participant who can perform tge verification, which limits its usefulness.
It is an extension of Tracker and focuses on improving performance and stricter path privacy requirements.
We did not notice any significant differences in terms of security between Cai et al.~\cite{caiNewFrameworkPrivacy2012} and Tracker~\cite{blassTrackerSecurityPrivacy2011}.

\textbf{Qian et al.}~\cite{qianLightweightRFIDSecurity2016a,qianLightweightPathAuthentication2018} focused on improving Tracker~\cite{blassTrackerSecurityPrivacy2011}.
Because their work mainly focuses on performance, the security they achieve is similar to that of Tracker.
Thus, they achieve path privacy and path authorization.
Like Tracker, it does not disclose the actual path to protect privacy.
Their scheme has the same security properties as Tracker~\cite{blassTrackerSecurityPrivacy2011} and Cai et al.~\cite{caiNewFrameworkPrivacy2012}.

\textbf{Checker}~\cite{elkhiyaouiCheckerOnsiteChecking2012} improves on Tracker~\cite{blassTrackerSecurityPrivacy2011} by allowing each reader to check whether a valid path has been taken.
Similar to PathChecker \cite{ouafiPathcheckerRFIDApplication2009}, they represented a path using a polynomial.
Each reader $r_i$ obtains its secret coefficient $a_i$.
They defined the tag state as a tuple $(Enc(ID)$, $Enc(path\_sig(ID))$, where $ID$ is the tag ID and $\sigma_p(ID)$ is the path signature of path $p$.
The path signature function is defined as $H(ID)^{Q_p(x_0)}$, where $H$ is a hash function and $Q_p(x_0)$ is the evaluation of the polynomial for path $p$ in point $x_0$.
Each reader $r_i$ has a list of paths $\{p_1, \dots, p_n\}$ that describe all valid paths ending in $r_i$.
Each reader $r_i$ also has a corresponding list of keys $\{K_1, \dots, K_n\}$ with which they can verify which valid path(if any) has been taken.

By allowing each reader to verify the path, they overcome the serious limitations of Tracker.
This also implies that their solution is \emph{sorted} under the $Adv_{T}$ adversary model.
However, they did not discuss the distribution of the valid paths.
Given that valid paths can change over time, there is also a significant overhead owing to the key distribution.

\textbf{Islam et al.}~\cite{islamIntegratingBlockchainSupply2022a} proposed a tracking system using RFID, a blockchain, and a physically uncloneable function. 
They also conducted an informal security analysis. 
Their threat model considered counterfeiting attacks (copying or reusing tags). 
They do not mention paths but their verification protocol verifies previous owners. 
Their approach does not consider path authorization or privacy requirements.

\textbf{PathChecker}~\cite{ouafiPathcheckerRFIDApplication2009} allows participants to track an RFID tag.
As reported by Ray et al.~\cite{raySecureObjectTracking2016} PathChecker relies on tamper resistant tags.
The tag must compute complex cryptography primitives, which limits the feasibility of such a system.
The authors assumed an honest reader adversary model and hard-coded the path into the tag secret.
This means that it satisfies our soundness and authorization properties.
PathChecker does not offer any privacy.

\textbf{Wang et al.}~\cite{wangTwolevelPathAuthentication2012} built a solution on top of the EPCGlobal network.
They split path authentication into two parts: within an EPCIS (organization) and between EPCISes. 
Within an EPCIS, traceability is achieved using hierarchical identity based encryption (HIBE) trees , batch verification is used between EPCISes.
Because of the two-level design, it did not fit our model perfectly. 
In our model, there is only one issuer. 
However, in this scheme, each EPCIS has an issuer.

A HIBE system is a generalization of identity-based encryption systems~\cite{bonehHierarchicalIdentityBased2005}.
It uses a tree to represent the hierarchy of identities within the system.
An identity at level $k$ can generate private keys for its descendants but cannot decrypt messages intended for other identities.
Therefore, Wang et al.~\cite{wangTwolevelPathAuthentication2012} created a hierarchy of readers within an EPCIS.
The initial tag secret is defined as $C_{{ID}_{R_i}} = (CT, \sigma^1)$, where $CT$ is the ciphertext of the tag ID and $\sigma^1$ is its signature.
Each subsequent step within an EPCIS updates $C_{{ID}_{R_i}} = (CT, \sigma^1)$ by encrypting their reader ID.

In their security analysis, they expected the path to be encoded on the tag.
After careful analysis, we could not determine how and where the path was encoded in this process.
HIBE encryption does not have any features that facilitate path storage.
In our view, the tag secrets updates without keeping track of its history.
Therefore we conclude that they do not provide path authentication or path authorization.
Because data are encrypted at all times, they provide a form of privacy.

They state that a reader $r$ might be involved in multiple paths.
The authors claim that ambiguity can be resolved by examining the tag signature.
However, this only works if the parent of reader $r$ is not involved in multiple paths.
For example consider two paths $t_1: r_1, r_2, r_3$ and $t_2: r_1, r_2, r_4$.
Once the tag arrives at $r_2$, the reader cannot know which reader is next based on the signature because both will have been signed by $r_1$.
Therefore, an additional context needs to be added to solve this problem.

Another problem is that the HIBE system has some inherent weaknesses~\cite{bonehHierarchicalIdentityBased2005}.
For example, tree depth degrades overall security.
They only achieve path authentication between organizations.
What happens within an organization is unknown to the other participants. 

Their solution requires $1600$ bits of storage.
However, this is beyond the scope of most RFID tags.

\textbf{DE-Sword}~\cite{qiDESwordIncentivizedVerifiable2022a} is a traceability querying system that works in a dishonest data owner model using a combination of zero-knowledge databases~\cite{zkdatabases} and game theory. 
Using incentives, they enforce that participants send correct data only. 
Each participant commits to the tags (products) it processed. 
When a participant gets queried, it responds with a proof of ownership or a proof of non-ownership. 
These proofs are unforgeable because he/she truthfully committed to all the processed tags.

Their system achieves \emph{path authentication} and \emph{path privacy} under the  $Adv_{R}$ adversary model. 
Currently, DE-Sword is vulnerable to replay and DoS attacks because secure communication protocols are absent. 
This vulnerability can easily be mitigated by adopting secure communication protocols. 
One limitation is that they assume a dishonest participant cannot distinguish a good product from a bad one.
They do not provide arguments as to why this would be the case.
A dishonest participant gains an advantage if he makes an oracle that can distinguish between good and bad products.
Another limitation is that querying must occur after distribution otherwise distinguishing between good and bad products is trivial.

\textbf{Zhang and Guin}~\cite{zhangEndtoEndTraceabilityICs2020} proposed a traceability solution for integrated circuit chips (ICs). 
More specifically, they presented an RFID-based solution that helps combat recycling attacks, where a dishonest participant sells old chips as new. 
Conventionally, every intermediate node on the path performs a re-measurement to verify the initial measurements.
Their approach only requires a re-measurement at the path end.
According to our analysis, they achieve path authentication.
However, they do not consider any other properties.
One of the main limitations is scalability.
The amount of tag memory required scales linearly with the number of steps in the path.
Because tag memory is small, the number of steps in a path must be relatively small.

\textbf{Cai et al.}~\cite{caiDistributedPathAuthentication2012a} proposed a path authentication scheme based on ordered multi-signatures, polynomial encoding, and Shamir secret-sharing.
An ordered multi-signature scheme allows participants to verify which participants signed the message in which sequence.
Their scheme operates with batches of tags instead of individual tags.
The tag contains a tag identifier $tid$, path code $pc$, and an ordered multi-signature $\sigma$.
The tag is encrypted with a key $k$.
This key is distributed to the tags using the Shamir secret-sharing algorithm.

A reader collects the key $k$ by collecting a sufficient number of shares.
Subsequently, it decrypts and reads the $tid$, $pc$, and signature $\sigma$.
Using an undisclosed communication protocol with the trusted server, it obtains the path and public keys by providing the path code $pc$.
Using this information, the reader can verify the signature and update the path code.
This does require the trusted server to store all sub-paths as well.

If an adversary only reads one tag, it cannot decrypt the content since it does not have enough shares to reconstruct the key.
However, it is reasonable to assume that if an adversary can read one tag, it can read all tags in the batch, which discloses the tag content and violates the privacy property.
The $\textsc{Path}(t, r_1, \dots, r_n)$ claim is reflected in two ways: in the signature $\sigma$ and the path code $pc$.
Since the multi-signature scheme is ordered it can only make valid traces:
\begin{multline}
tr_{valid} = \textsc{Move}(t, r_1) \rightarrow \dots \rightarrow \textsc{Move}(t, r_n) \rightarrow \\
\textsc{Path}(t, r_1, \dots, r_n)
\end{multline}
Because the path code and the signature have to match, we know that $\textsc{Path}(t, r_1, \dots, r_n)$ must be one of the valid paths.
Therefore, their protocol satisfies both path authentication and authorization.
In our adversary model, path privacy is not satisfied because the adversary can reconstruct the key $k$.

\subsection{Key takeaways}
\label{sec:discussion}
During our analysis, we observed that most traceability solutions focus on achieving \emph{path soundness} and \emph{path privacy}. \emph{Path completeness} and \emph{authorization} are often not discussed.
On the one hand, 
\emph{path completeness} is essential in situations with passive adversaries and can prevent reroute attacks.
For example, consider a product is intercepted between $r_i$ and $r_j$ when a malicious reader $r_x$ scans the product without changing anything.
The product is tampered with before $r_x$ sends it to $r_j$.
However, path completeness seems difficult to achieve, as it is hard to monitor all movements of a tag without making strong assumptions about the tags, such as the presence of a GPS receiver. 

Path authorization has mostly been neglected in the literature despite its practical significance in preventing supply chain management errors. 
Studies that provide path authorization require a secure distribution of valid paths or policies~\cite{qianLightweightPathAuthentication2018,wangEfficientTagPath2016,blassTrackerSecurityPrivacy2011,burbridgeSupplyChainControl2010,raySecureObjectTracking2016}.
None of these studies described the establishment or distribution of these valid paths.
We hypothesize that this prerequisite for some traceability systems might become a popular target for attack.
If an adversary can influence policies, there is no reason to attack the traceability system.
We believe that this aspect of traceability systems should receive more attention in the scientific literature.

We noticed that some solutions~\cite{blassTrackerSecurityPrivacy2011,ouafiPathcheckerRFIDApplication2009} do not authenticate a single path but merely require that a valid path from a set of valid paths is taken.
Unless the valid path is unique, path authentication cannot be supported.
Some solutions, such as in Wang et al. ~\cite{wangTwolevelPathAuthentication2012}, can authenticate that a certain path has been taken without guaranteeing the path's validity.
Hence, we divide the group of systems into two categories. 
The first category allows readers to retrieve the traveled path (explicitly), whereas the second category shows that a path from a given set of paths is followed.

We encourage new traceability solutions to examine these issues.
Creating a traceability solution that provides complete path authentication and can prove that its policies are established openly and honestly, would greatly enhance the security of the overall system.

Finally, our analysis approach differs from existing approaches that aim to capture data, such as sensor readings and RFID reads/writes, and store them on a secure platform.
In this case, security means ensuring that the data are accurate, valid, complete, and consistent~\cite{gs1GS1GlobalTraceability}.
However, our framework challenges this approach.
Using path-based security properties, we can define what a secure traceability system aims to achieve with precision.
We also define all the types of attack, outlined by our attack taxonomy in Section \ref{sec:framework-attacks}.
By applying our methodology we found several weaknesses and vulnerabilities in recent papers, supporting the benefits of the path-based framework introduced in Section \ref{sec:framework}.
We found vulnerabilities in ReSC~\cite{yangReSCRFIDEnabledSolution2018}, RF-Chain~\cite{wangRFChainDecentralizedCredible2023}, Ray et al.~\cite{raySecureObjectTracking2016}, and Burbridge et al.~\cite{burbridgeSupplyChainControl2010}. RF-Chain contains a linking attack that invalidates our privacy property.
Ray et al. do not require the readers to be followed in order. 
Similarly, in Tracker~\cite{blassTrackerSecurityPrivacy2011} readers do not need to be followed in order because of the commutative property of polynomials.
The work by Burbridge and Soppera~\cite{burbridgeSupplyChainControl2010} includes an attack on the \emph{authorization} property in the $Adv_{R}$ adversary model.
We also found weaknesses in numerous traceability systems, such as DE-Sword~\cite{qiDESwordIncentivizedVerifiable2022a}, SecTTS~\cite{Shi2012}, and SupAuth~\cite{mamunSupAUTHNewApproach2018}.
DE-Sword was challenging to analyze within our framework, implying that we need to extend our framework in future work.
\subsection{Limitations}
The main limitation of our analysis methodology and framework is that it is limited to a guided, yet semi-informal, analysis. 
However, the goal of analyzing traceability systems with full mathematical rigor on a large scale seems unattainable in the near future because of the large variety of design approaches and absence of verification tools that deal well with physical properties.
A limitation of the framework, not the methodology, is its inability to formally capture the privacy properties of interest.  
These issues will be addressed in future studies.

\section{Conclusion}
\label{sec:conclusion}
This study introduces a novel evaluation framework for the security of RFID-enabled traceability systems for supply chains. 
The framework is generic enough to capture most traceability systems in the literature, yet sufficiently detailed to provide relevant results which we have shown in various case studies. 
It does so by using paths as building blocks for our security properties, offering a new perspective on the security analysis of traceability systems. 
In particular, by applying our methodology and framework, we identified several weaknesses and vulnerabilities in the existing work.
Our study also provides a taxonomy for path-based attacks on traceability systems.

There are several directions for future research. 
First, the scope can be extended to traceability solutions that rely on other tracking technologies (besides RFID), such as GPS transmitters and IoT sensors.
Second, we believe that with some effort these traceability systems can be modeled using the verification tool \textsc{Tamarin}~\cite{basinTamarinVerificationLargeScale2022}, enabling formal verification and thus reaching full mathematical rigor. 
Third, we seek solutions to prevent reroute attacks in physical supply chains, a type of attack popular in routing protocols but less explored in the supply chain. 
Fourth, several schemes seem to allow for out-of-order attacks. Future research should look deeper into this and fix these schemes such that they are resistant to this type of attack.
Finally, our work can be extended by including security and privacy properties that are not path-based.


 \section*{Acknowledgements}
The work has been supported by the Cyber Security Research Centre Limited whose activities are partially funded by the Australian Government’s Cooperative Research Centres Programme.

Rolando Trujillo-Rasua is funded by a Ramon y Cajal grant from the Spanish Ministry of Science and Innovation and the European Union (REF: RYC2020-028954-I). 
He is also supported by the project HERMES funded by INCIBE and the European Union NextGenerationEU/PRTR.

\bibliographystyle{IEEEtran}  
\bibliography{sources}  

\end{document}